\documentclass{article}
\usepackage{epsfig,rotate,multicol,graphicx,psfrag,bm,times}
\begin{document}

\title{Running coupling constant from lattice data and bulk viscosity of strange quark matter }
\author{Zheng Xiaoping, Kang Miao, Liu Xuewen, Yang Shuhua
\\
{\small  The Institute of Astrophysics, Huazhong Normal
University, Wuhan 430079 P. R. China }}
\date{}
\maketitle
\begin{abstract}
  We study the bulk viscosity of strange quark
  matter (SQM) in a quasiparticle model at finite chemical potential by extrapolating the previous quasiparticle model of finite
   temperature lattice QCD. The more proper bulk viscosity coefficient
   can be given in this  model where chemical
   potential $\mu$ and coupling constant $\emph{g}$ are interdependent.
   We also apply our result to determine the critical rotation
   of strange stars by r-mode instability window. Our model is
   compatible to the millisecond pulsar data for a wide range of
   mass and radius of the stars.
  \\
PACS numbers: 97.60.Jd, 12.38.Mh, 97.60.Gb
\end{abstract}
\section{INTRODUCTION}

$\indent$ According to Witten's conjecture that the strange quark
matter (SQM) composed of roughly equal numbers of up, down and
strange quarks might be absolutely stable or metastable phase of
nuclear matter in 1984 \cite{1}, strange stars might exist in the
universe, and their structure and properties have been widely
studied \cite{2,3,4,5,6,7}.

How to distinguish strange stars from neutron stars has been one
of the important issues. Madsen pointed out that the r-mode
instability in the relativistic stars at all rotating rates may
provide a distinguishable signature \cite{8}. The main reason is
that the bulk viscosity coefficient of SQM is larger than that of
neutron matter, and hence suppresses the r-modes. Since Wang and
Lu \cite{9} found that the non-leptonic weak reaction dominates
the bulk viscosity of SQM, some investigations have tried to
calculate the relevant viscosity coefficient of SQM. The equation
of state (EOS) of strange stars has also been studied as the base
of studying the viscosity. Sawyer \cite{2} and Madsen \cite{3}
completed the calculations of bulk viscosity for the ideal quark
gas.  We had investigated  the viscosity of interacting SQM in
quasiparticle description by regarding the coupling constant as an
independent parameter and found the improved result is of
importance for astrophysical relevance \cite{10}. However it is
desirable to have a more consistent investigation by considering
the realistic running coupling constant for SQM in the interior of
compact stars. According to the regulations, the realistic
coupling constant and EOS need nonperturbative evaluations,
i.~e.~lattice QCD calculations. However, the present lattice QCD
calculations are yet restricted to zero chemical potential $\mu$.
Attempts to extend the lattice calculations systematically to
non-zero chemical potential are underway. Being aware of the
urgent need for the EOS, many works have suggested an approach
based on a quasiparticle description of quarks and gluons, to map
available lattice data from $\mu=0$ to finite values of $\mu$ and
to small temperatures \cite{11}. Consequently, the running
coupling in the asymptotic limit of large chemical potential can
be also simulated from that of large temperatures at zero chemical
potential, which will also help us to overcome the difficulty of
calculating EOS in non-perturbation regime. We thus can
consistently evaluate the bulk viscosity in the light of the
simulated EOS and running coupling constant. We will see below the
coupling among quarks  influences remarkably the bulk viscosity in
SQM. Our theoretical output is  appropriate  for a wide range of
stellar parameters.

     We organize this paper as follows. In Sec.2, we introduce the
 equation of state and formulate the quasiparticle model at finite chemical.
  In Sec.3, the bulk viscosity of SQM in the quasiparticle
  model is derived, which arises from the nonleptonic weak
  interaction.  In Sec.4, we probe the application of our model. Finally, we summarize our conclusion and
  discussion in Sec.5.

\section{THE QUASI-PARTICLE MODEL}

 $\indent$ Matter in local thermodynamical equilibrium can be described by
its EOS which represents an important interrelation of state
variables. For the EOS of SQM, nonperturbative methods as lattice
QCD should be applied. As a matter of fact, these simulations are
presently restricted to vanishing chemical potential $\mu$, and
the implementation of physical quark masses is still too expensive
numerically. So we often consider phenomenological model. The bag
model EOS is popular, but it is in conflict with thermodynamic QCD
lattice data. We will apply the quasipaticle model and  here
repeat a simple extrapolation of finite temperature lattice to
nonzero chemical potential relying on thermodynamic consistency
\cite{11}.

 Asymptotically, the collective behavior of the plasma can be
 interpreted in terms of quasiparticle excitations with a
 dispersion relation $\omega_{i}^{2}(\emph{k})\approx\emph{m}_{i}^{2}+\emph{k}^{2}$
 and $\emph{m}_{i}^{2}=\emph{m}_{0i}^{2}+\Pi_{i}^{*}$ depending on
 the rest mass and the leading order on-shell
 self-energies \cite{12},
\begin{equation}
\Pi_{q}^{*}=2\omega_{q0}(\emph{m}_{0}+\omega_{q0}),~~~~~~
\omega_{q0}^{2}=\frac{\emph{N}_{c}^{2}-1}{16\emph{N}_{c}}\left[\emph{T}^{2}+\frac{\mu_{q}^{2}}{\pi^{2}}\right]\emph{g}^{2}\\,
\end{equation}
\begin{equation}
\Pi_{g}^{*}=\frac{1}{6}\left[(\emph{N}_{c}+\frac{1}{2}\emph{N}_{f})\emph{T}^{2}+\frac{3}{2\pi^{2}}\sum_{q}\mu_{q}^{2}\right]\emph{g}^{2},
\end{equation}
where $\mu_{q}$ denotes the quark chemical potential, and
$\emph{N}_{c}$=3. The pressure of quasiparticle system can be
decomposed into the contributions of the quasiparticles and their
mean-field interaction $\emph{B}$,
\begin{equation}
\emph{p}(\emph{T},\mu;\emph{m}_{0j}^{2})=\sum_{i}\emph{p}_{i}[\emph{T},\mu_{i}(\mu);\emph{m}_{i}^{2}]-\emph{B}(\Pi_{j}^{*}),
\end{equation}
where
$\emph{p}_{i}=\pm\emph{d}_{i}\emph{T}\int\emph{d}^{3}\emph{k}/(2\pi)^{3}\ln(1\pm\exp\{-(\omega_{i}-\mu_{i})/\emph{T}\})$
are the contributions of the gluons (with vanishing chemical
potential) and the quarks (for the anti-quarks, the chemical
potential differs in the sign). And
$\emph{d}_{g}=2(\emph{N}_{c}^{2}-1)$ and
$\emph{d}_{q}=2\emph{N}_{c}$ count the degrees of freedom
\cite{13}. The function $\emph{B}(\Pi_{j}^{*})$ is determined from
a thermodynamical self-consistency condition, via
\begin{equation}
\frac{\partial\emph{B}}{\partial\Pi_{j}^{*}}=\frac{\partial\emph{p}_{j}(\emph{T},\mu_{j};m_{j}^{2})}{\partial\emph{m}_{j}^{2}}.
\end{equation}
Furthermore, the stationarity implies that the entropy and the
particle densities are given by the sum of the quasiparticle
contributions,
\begin{equation}
\emph{s}_{i}=\frac{\partial\emph{p}_{i}(\emph{T},\mu_{i};m_{i}^{2})}{\partial\emph{T}}\bigg|_{\emph{m}_{i}^{2}},~~~~~~
\emph{n}_{i}=\frac{\partial\emph{p}_{i}(\emph{T},\mu_{i};m_{i}^{2})}{\partial\mu_{i}}\bigg|_{\emph{m}_{i}^{2}},
\end{equation}
while the energy density has the form
$\emph{e}=\sum_{i}\emph{e}_{i}+\emph{B}$. By comparison with
lattice data, at $\mu=0$, Peshier\cite{11} has tested the
quasiparticle approach which is an appropriate description even
close to the confinement transition, with the effective coupling
in Eq (1) and (2) nonperturbatively parameterized by
\begin{equation}
\emph{g}^{2}(\emph{T},\mu=0)=\frac{48\pi^{2}}{(11\emph{N}_{c}-2\emph{N}_{f})\ln\left(\frac{T+T_{s}}{T_{c}/\lambda}\right)^{2}},
\end{equation}
interpolating to the asymptotic limit of QCD.

Encouraged by the successful quasiparticle description of the
$\mu=0$ lattice data, the model is now extrapolated to finite
chemical potential. In general, the pressure is a function of the
state variables $\emph{T}$ and $\mu$. As a direct consequence
thereof, the Maxwell relation implies for the quasiparticle model
\begin{equation}
\sum_{i}\left[\frac{\partial\emph{n}_{i}}{\partial\emph{m}_{i}^{2}}\frac{\partial\Pi_{i}^{*}}{\partial\emph{T}}-\frac{\partial\emph{s}_{i}}{\partial\emph{m}_{i}^{2}}\frac{\Pi_{i}^{*}}{\mu}\right]=0,
\end{equation}
which is the integrability condition for the function $\emph{B}$
defined by Eq (4). Following directly from principles of
thermodynamics, we can get a flow equation for the effective
coupling with $\Pi_{i}^{*}$ depending on $\emph{g}^{2}$. This flow
equation is a quasilinear partial differential equation of the
form
\begin{equation}
\emph{a}_{T}\frac{\partial\emph{g}^{2}}{\partial\emph{T}}+\emph{a}_{\mu}\frac{\partial\emph{g}^{2}}{\partial\mu}=\emph{b},
\end{equation}
with the coefficients $\emph{a}_{T,\mu}$ and $\emph{b}$ depending
on $\emph{T}$, $\mu$ and $\emph{g}^{2}$.

The flow of the effective coupling is elliptic in the
nonperturbative regime, thus mapping the $\mu=0$ axis, where
$\emph{g}^{2}$ can be determined from lattice data, into the
$\mu-\emph{T}$ plane. As an example, the nonperturbative flow of
the coupling of the $\emph{N}_{f}=2+1$ system is shown in
Figure.1. Here we use the parameters $\lambda=6.6$ and
$\emph{T}_{s}=-0.78\emph{T}_{c}$ \cite{14}. Furthermore, the
running coupling constant
 as a function of chemical potential $\emph{g}(\emph{T}=0,\mu)$ is displayed in
Figure.2. For convenience later, we can approximately formulate
$\emph{g}(\emph{T}=0,\mu)$ as
\begin{equation}
\emph{g}^{2}(\emph{T}=0,\mu)=\frac{48\pi^{2}}{(11\emph{N}_{c}-2\emph{N}_{f})\ln(\frac{\mu+T_{s}}{T_{c}\pi/\lambda})^{2}}.
\end{equation}
Based on the extended coupling constant, the EOS of SQM in the
interior of compact stars is immediately obtained from Eqs (1),
(2), (3) and (4).

\begin{figure}[ht]
\centerline{\epsfig{file=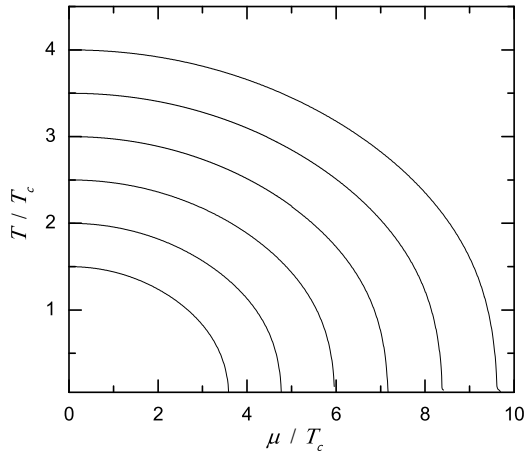,  width=8.6cm}} \caption{The
characteristics of the coupling flow equation (8) for the QCD
plasma with $\emph{N}_{f}=2+1$ flavors.}\label{mfig1}
\end{figure}
\begin{figure}[ht]
\centerline{\epsfig{file=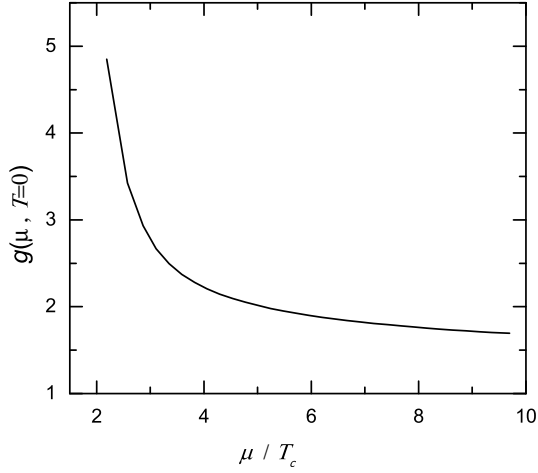,  width=8.6cm}} \caption{The
coupling constant $\emph{g}(\emph{T}=0,\mu)$ as a function of
$\mu$ .}\label{mfig2}
\end{figure}

\section{BULK VISCOSITY}

$\indent$Viscosity is important for describing the transport
property of matter and can be generally calculated from either
quantum field theory or kinetic theory\cite{15}. The bulk
viscosity of SQM mainly arises from the nonleptonic weak
interactions \cite{9}
\begin{equation}
\emph{u}+\emph{d}\longleftrightarrow\emph{s}+\emph{u}.
\end{equation}
The importance of dissipation due to the reaction (10) was first
stressed by Wang and Lu in the case of neutron stars with quark
cores. Afterward, a series of investigations have tried to
calculate the viscosity coefficient of SQM in MIT bag model and
showed a huge bulk viscosity of strange matter relative to nuclear
matter. Furthermore, Zheng et al \cite{10} considered the medium
effect on the bulk viscosity in quasiparticle description and
found the viscosity is few$\sim$tens times larger than that of
non-interacting quark gas due to the small influence of medium on
the equation of state, where coupling constant $\emph{g}$ is
regarded as an independent parameter of chemical potential. In
fact, $\mu$ and $\emph{g}$ should be interdependent according to
the results of lattice calculation. We here reevaluate the bulk
viscosity coefficient adopting the description in Sec.2.

 In quasiparticle approximation, we can give the formulae of the bulk viscosity
 in accordance with Eqs (1), (2), (3) and the reactional rate of reaction (10) at finite chemical potential when temperature
 is small. Zheng et al \cite{10} have made derivations
of bulk viscosity in quasiparticle description. We can apply the
result,
\begin{equation}
\zeta=\frac{1}{\pi\upsilon_{0}}\left(\frac{\upsilon_{0}}{\Delta\upsilon}\right)\frac{1}{3}
\left(\frac{k_{Fd}^{2}}{C_{d}}-\frac{k_{Fs}^{2}}{C_{s}}\right)\int_{\tau}^{0}d\emph{t}
\left[\int_{0}^{t}\frac{dn_{d}}{dt}\right]\cos\left(\frac{2\pi\emph{t}}{\tau}\right),
\end{equation}
where $\emph{k}_{Fi}=(\mu_{i}^{2}-\emph{m}_{i}^{2})^{1/2}$,
$\emph{C}_{i}=\mu_{i}-\sqrt{\frac{1}{6}}\frac{\emph{g}{\emph{m}}_{0}}{\pi}
-\frac{1}{3}\frac{\mu}{\pi^{2}}{\emph{g}}^{2}-\sqrt{\frac{1}{6}}\frac{\mu{\emph{m}}_{0}}{\pi}\frac{\partial\emph{g}}{\partial\mu}
-\frac{1}{6}\frac{\mu^{2}}{\pi^{2}}\frac{\partial\emph{g}^{2}}{\partial\mu}$,
the expression of $\emph{C}_{i}$ has been reevaluated, differing
from that of Zheng et al \cite{10} because $\emph{g}$ is the
function of $\mu$. The effective mass $\emph{m}_{i}$ was given in
Sec.2, the current masses vanish for up and down quarks. We
continue to use the formula of the reactional rate adopted by
Madsen \cite{3}
\begin{equation}
\frac{dn_{d}}{dt}\approx\frac{16}{5\pi^{2}}G_{F}^{2}\sin^{2}\theta_{c}\cos^{2}\theta_{c}\mu_{d}^{5}\delta\mu
[\delta\mu^{2}+4\pi^{2}T^{2}]\upsilon_{0},
\end{equation}
with \cite{10}
\begin{equation}
\delta\mu=\frac{1}{3}\left(\frac{k_{Fd}^{2}}{\emph{C}_{d}}-\frac{k_{Fs}^{2}}{\emph{C}_{s}}\right)
\frac{\triangle\upsilon}{\upsilon}\sin\left(\frac{2\pi\emph{t}}{\tau}\right)-\frac{\pi^{2}}{3}\frac{1}{\upsilon}
\left(\frac{1}{k_{Fd}C_{d}}-\frac{1}{k_{Fs}C_{s}}\right)\int_{0}^{t}\frac{dn_{d}}{dt}.
\end{equation}

%From these equation, we can get the results numerically of bulk
%viscosity coefficient.

When the temperature is high enough, i.~e.
$2\pi\emph{T}\gg\delta\mu$, the cubic term $\delta\mu^{3}$ in Eq
(12) can be neglected. We can obtain an analytical result which is
similar to Madsen's expression \cite{3},
\begin{equation}
\zeta=\frac{\alpha^{*}T^{2}}{\omega^{2}+\beta^{*}T^{4}}
\end{equation}
and
\begin{equation}
\alpha^{*}=9.39\times10^{22}\mu_{d}^{5}\left(\frac{k_{Fd}^{2}}{C_{d}}-\frac{k_{Fs}^{2}}{C_{s}}\right)^{2}
(g~cm^{-1}s^{-1}),
\end{equation}
\begin{equation}
\beta^{*}=7.11\times14^{-4}\left[\frac{\mu_{d}^{5}}{2}\left(\frac{1}{k_{Fd}C_{d}}-\frac{1}{k_{Fs}C_{s}}\right)\right]^{2}(s^{-2}),
\end{equation}
here $\omega=\frac{2\pi}{\tau}$, $\alpha^{*}$ and $\beta^{*}$
remarkably differ from Madsen's.

In general, Eqs (11), (12) and (13) must be solved numerically due
to the existence of $\delta\mu^{3}$ in rate (12). The results of
such calculations are shown in Figure 3 and Figure 4. The
magnitude of viscosity coefficient increases with chemical
potential for the low-temperature case such as $\emph{T}=10^{-4}$
MeV in Figure 3. It is just contrary to the high-temperature case
as $\emph{T}=10^{-1}$ MeV in Figure 3. This implies a shift of the
the maximal viscosity to the low-temperature when chemical
potential increases as shown in Figure 4. By comparison, the
viscosity has more complicated dependence on coupling constant
than that of ref \cite{10}.

\begin{figure}[ht]
\centerline{\epsfig{file=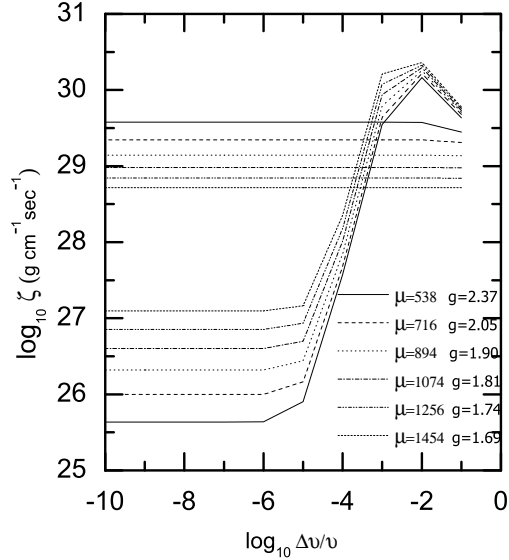,  width=8.6cm}} \caption{Bulk
viscosity coefficient of different $\mu$ and $\emph{T}$,
$\emph{m}_{0s}$=200 MeV, $\tau$=0.001 s. The lower curves denote
$\emph{T}=10^{-4}$ MeV, the upper curves denote $\emph{T}=10^{-1}$
MeV. }\label{mfig3}
\end{figure}
\begin{figure}[ht]
\centerline{\epsfig{file=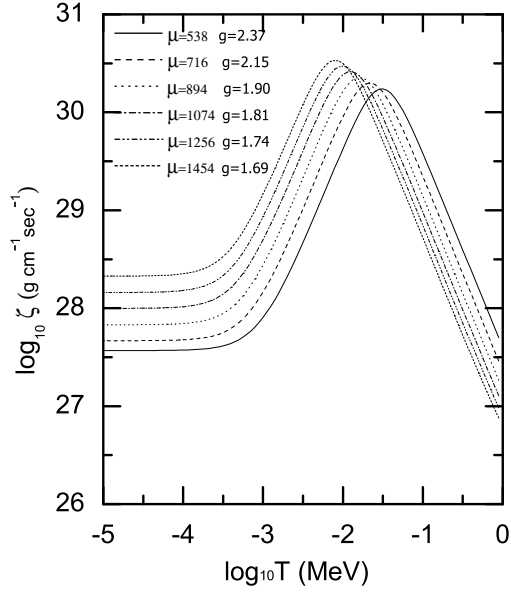,  width=8.6cm}} \caption{Bulk
viscosity as function of temperature for different $\mu$,
$\emph{m}_{0s}$=200 MeV, $\tau$=0.001 s,
$\frac{\Delta\upsilon}{\upsilon}=10^{-4}$.}\label{mfig4}
\end{figure}

\section{CRITICAL ROTATION OF STRANGE STARS}
$\indent$Given the EOS of SQM, we can get radius-mass
characteristic relationship of strange stars through the
Tolman-Oppenheimer-Volkov(TOV) equation \cite{16}
 for energy densities up to several times the nuclear saturation
 density and at temperatures less than some 10 MeV, the relation
 $\emph{e}(\emph{p})$ of $\beta$-stable quark matter can be
 parametrized\cite{11} by $\emph{e}=4\tilde{\emph{B}}+\tilde{\alpha}\emph{p}$, as
 estimated numerically by our quasi-particle model, here energy density $\emph{e}=
 \emph{e}_{u}+\emph{e}_{d}+\emph{e}_{s}+\emph{e}_{e}+\emph{B}_{0}$ and
 pressure $\emph{p}=\emph{p}_{u}+\emph{p}_{d}+\emph{p}_{s}+\emph{p}_{e}-\emph{B}_{0}$. For the
 chosen parameters in Sec.2 and the bag constant $\emph{B}_{0}^{1/4}=145$ Mev, we can get $\tilde{\alpha}=3.5$ and
$\tilde{\emph{B}}^{1/4}=152$ MeV in our EOS.
 Furthermore, we can find corresponding values of the parameter
 $\tilde{\emph{B}}$ by changing the bag constant
 $\emph{B}_{0}$. The values of $\tilde{\emph{B}}$ have a strong impact
 on the star's mass and radius, obtained by integrating the TOV
 equation. In Figure.5, the mass as a function of the radius of strange stars
is displayed for several values of the parameter
$\tilde{\emph{B}}$. The maximal mass and radius are essentially
consistent with canonical pulsar data for
$\emph{B}_{0}^{1/4}\sim145$ Mev and
$\tilde{\emph{B}}^{1/4}\sim152$ MeV.

Now we focus on the r-mode unstable window for this case.
Andersson in 1998 \cite{17} recognized the existence of unstable
r-modes in perfect fluid stars at all rates of rotation due to
gravitational wave emission. The r-mode unstable regime of the
realistic
 stars, neutron stars as well as strange stars, depends on the
 competition between the gravitational radiation and various
 dissipation mechanisms. To plot the instability window of r-mode
 or obtain the critical rotation frequency for given stellar model, we need to work out the
 characteristic timescales,damping and growing timescales of
 r-mode instability. The time scale for gravity wave emission is \cite{18}
\begin{equation}
\tau_{G}=\overline{\tau}_{G}(\pi\emph{G}\overline{\rho}/\Omega^{2})^{3},
\end{equation}
where, $\overline{\tau}_{G}$ is -3.26 s for $n=1$ polytropic EOS,
$\overline{\rho}$ is the mean density of the star. In strange
stars, the time scales for shear and bulk viscous dissipations can
be respectively given
\begin{equation}
\tau_{S}=\overline{\tau}_{S}(\alpha_{s}/0.1)^{5/3}T_{9}^{5/3},
\end{equation}
\begin{equation}
\tau_{B}=\overline{\tau}_{BL}(\pi\emph{G}\overline{\rho}/\Omega^{2})T_{9}^{-2}
\end{equation}
for low-$\emph{T}$ limit,
\begin{equation}
\tau_{B}=\overline{\tau}_{BH}(\pi\emph{G}\overline{\rho}/\Omega^{2})^{2}T_{9}^{2}
\end{equation}
for high-$\emph{T}$ limit, where $\overline{\tau}_{s}$ is
5.37$\times10^{8}$ s for n=1,
$\overline{\tau}_{BL}=2.83\times10^{3}\alpha^{*-1}\overline{\rho}m_{100}^{4}$
\cite{6,7}, $\alpha_{s}=\frac{g^{2}}{4\pi}$ is QCD fine structure
constant, $\emph{T}_{9}$ denotes temperature in units of $10^{9}$
K, $m_{100}$ denotes the current mass of strange quark in units of
100 MeV. Obviously, $\overline{\tau}_{B}$ is determined with the
chemical potential $\mu_{d}$. For given mean density
$\overline{\rho}$, We can find appropriate $\mu_{d}$ and
$\emph{g}$ via conservation of baryon number,
\begin{equation}
n_{B}=\frac{1}{3}(n_{u}+n_{d}+n_{s})
\end{equation}
and
\begin{equation}
\overline{\rho}=\left(\frac{E}{A}\right)n_{B},
\end{equation}
here, $n_{B}$ reprensents the baryon number density, and
$\emph{n}_{i}=\frac{1}{6\pi^{2}}\emph{k}_{Fi}^{2}$ is the flavor
number density, $\frac{E}{A}=\frac{
\emph{e}_{u}+\emph{e}_{d}+\emph{e}_{s}+\emph{e}_{e}+\emph{B}_{0}}{n_{B}}$
is the energy per baryon.

 We can evaluate the the critical spin frequency as a function of
temperature from the equation
\begin{equation}
\frac{1}{\tau_{G}}+\frac{1}{\tau_{S}}+\frac{1}{\tau_{B}}=0.
\end{equation}

 Figure.6 shows the regions of r-mode (in)stability in spin
 frequency-temperature($\nu-\emph{T}$) plane for a strange star
 with mass $\emph{M}=1.4~\emph{M}_{\odot}$ and radius
 $\emph{R}=10$ km. In comparision, the instability window of usual strange stars is also depicted in this figure,
  which is accomplished by Madsen. We find the medium effect narrows the
 r-mode instability window. The lowest limiting frequency of upper contour is
 541 Hz (the corresponding period is 1.85 msec), which is more close to
 the most rapidly spinning pulsars with frequencies of 642 Hz and
 622 Hz (the periods are 1.56 and 1.61msec). This implies a strange star would slow down by
 gravitational window and spin around in 1.85 msec instead of the
 $2.5\sim3$ msec expected by Madsen. Figure.7 describes the instability
 windows for a wide parameters $\tilde{\alpha}$, $\tilde{\emph{B}}$.
 The dotted curves denote a strange star with mass $\emph{M}=1.0~\emph{M}_{\odot}$ and radius $\emph{R}=6.6$ km,
 and the solid curves express a strange star with mass $\emph{M}=2.06~\emph{M}_{\odot}$ and radius $\emph{R}=13.5$ km.

\section{SUMMARY AND DISCUSSION}
$\indent$In accordance with the quasiparticle description, we
rederive and reevaluate the bulk viscous coefficient of SQM basing
on referring $\mu$ dependence of the coupling constant. We find
that below few times $10^{9}$ K, the medium effect is enhancing
the viscosity. The dissipation of fluctuations in dense matter is
more effective for this situation. We apply our result to study
the r-mode instability window of strange stars. We find the r-mode
instability window is narrowed and the lowest limiting frequency
is closer to the millisecond pulsars relative to Madsen's one for
stars with mass $\emph{M}=1.4~\emph{M}_{\odot}$ and radius
$\emph{R}=10$ km. Our model also allow for wide frequency changes
from 400 Hz to 900 Hz. The scenario is compatible with pulsar data
for a wide range of mass ($1.0\sim2.8~\emph{M}_{\odot}$) and
radius ($6\sim15$ km).

Acknowledgements: This research was supported by NSFC grant Nos
10373007 and 90303007, and the Ministry of Education of China with
project No 704035. We would like to thank Professor Li Jiarong for
the useful discussion.

\begin{figure}[ht]
\centerline{\epsfig{file=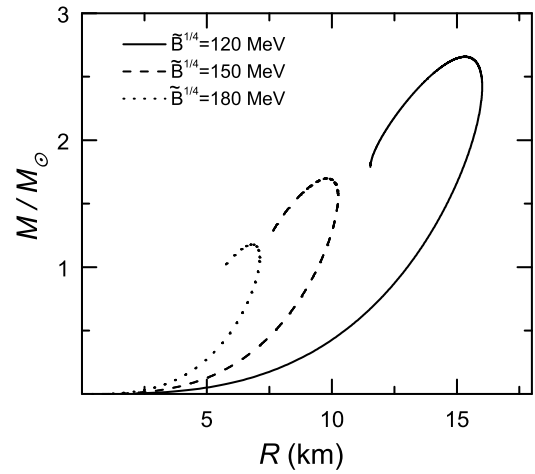,  width=8.6cm}} \caption{The
dependence of the mass of strange stars on the radius for
$\tilde{\alpha}=3.5$ and several values of parameters
$\tilde{\emph{B}}^{1/4}$.}\label{mfig5}
\end{figure}
\begin{figure}[ht]
\centerline{\epsfig{file=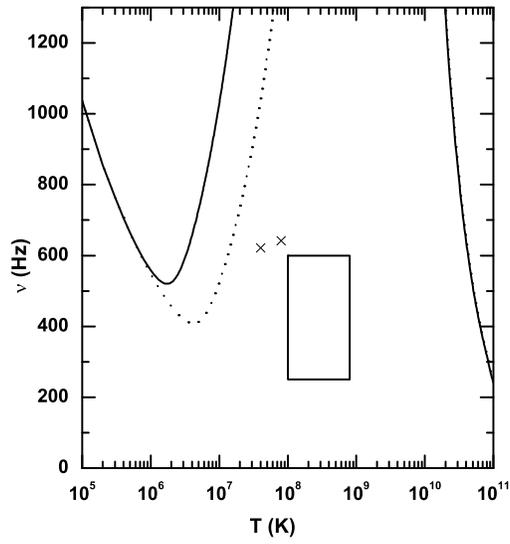,  width=8.6cm}}
\caption{Critical spin frequencies for strange stars as functions
of temperature with $\emph{M}=1.4\emph{M}_{\odot}$ and
$\emph{R}=10$km. The dotted contour stand for Madsen's model,
 the solid curves display our result.}\label{mfig6}
\end{figure}
\begin{figure}[ht]
\centerline{\epsfig{file=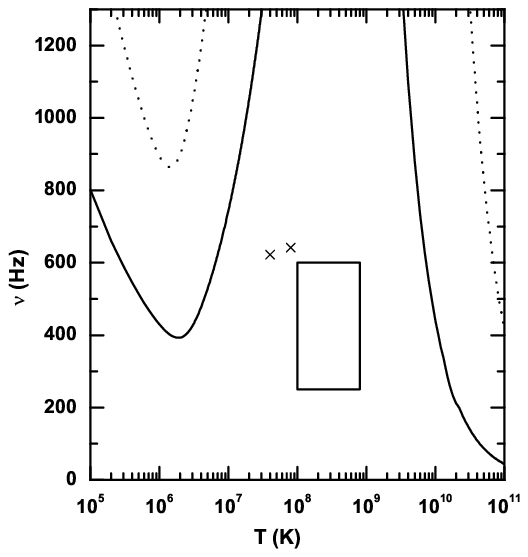,  width=8.6cm}}
\caption{Critical spin frequencies for strange stars as functions
of temperature. The dotted contour stand for a strange star with
$\emph{M}= 1.0\emph{M}_{\odot}$ and $\emph{R}=6.6$km,
 the solid curves display
a strange star with $\emph{M}= 2.06\emph{M}_{\odot}$ and
$\emph{R}=13.5$km.}\label{mfig7}
\end{figure}

\begin{thebibliography}{99}
\bibitem{1}E. Witten, Phys. Rev. D 30, 272 (1984).
\bibitem{2}R. F. Sawyer, Phys. Lett. B 233, 412 (1989).
\bibitem{3}J. Madsen, Phys. Rev. D 46, 3290 (1992).
\bibitem{4}A. Goyal, V. K. Gupta. and J. D. Anand, Z. Phys. A 349, 93 (1994).
\bibitem{5}k. Schertler,  C. Greiner and M. H. Thoma, Nucl. Phys.
A 616, 659 (1997).
\bibitem{6}J. Madsen, Phys. Rev. Lett. 85, 10 (2000).
\bibitem{7}X. P. Zheng, S. H. Yang, J. R. Li,
Astrophys. J. Lett. 585, L135 (2003).
\bibitem{8}J. Madsen, Phys. Rev. Lett. 81, 3311 (1998).
\bibitem{9}Q. D. Wang, T. Lu, Phys. lett. B 148, 211 (1984).
%\bibitem{10}G. Baym, S.A. Chin, Nucl. Phys. A262(1976)527.
%\bibitem{11}B. Freedman, Mclerran, Phys. Rev. D17(1978)1109.
%\bibitem{12}S. Chakrabarty, S. Raha, B. Sinba, Phys. Lett. B229(1989)112.
%\bibitem{13}S. Chakrabarty, Phys. Rev. D48(1993)1409.
%\bibitem{14}G. X. Peng, H. C. Chiang, P. Z. Ning, Phys. Rev.
%C62(1997).
\bibitem{10} X. P. Zheng, S. H. Yang, J. R. Li and X. Cai, Phys. Lett.
B 548, 29 (2002).
\bibitem{11}A. Peshier, B. Kampfer, and G. Soff, Phys. Rev.
C 61, 045203 (2000).
\bibitem{12}M. Le Bellac, $\emph{Thermal Field Theory}$ (Cambridge
University Press, Cambridge, 1996).
\bibitem{13}A. Peshier, B. Kampfer, and G. Soff, Phys. Rev.
D 66, 094003 (2002).
\bibitem{14}A. Peshier, B. Kampfer, and G. Soff, hep-ph/0106090.
\bibitem{15}M. E. Carrington, D. F. Hou, R. Kobes, Phys. Rev. D 64, 025001
(2003). ibid, 62,025010(2000).
\bibitem{16}N. K. Glendenning, $\emph{Compact Stars}$ (Springer-Verlag, New york,
1997).
\bibitem{17}N. Andersson, Astrophys. J. 502, 708 (1998).
\bibitem{18}L. Lindblom, G. Mendel, B. J. Owen, Phys. Rev.
D 60, 064007 (1999).
\end{thebibliography}
\end{document}